# Luttinger Liquid Behavior in Carbon Nanotubes

Marc Bockrath*, David H. Cobden*, Jia Lu*, Andrew G. Rinzler[+], Richard E. Smalley[+], Leon Balents[#]

and Paul L. McEuen*

*Department of Physics, University of California and Materials Sciences Division, Lawrence Berkeley National Laboratory, Berkeley, California, 94720

[+]Center for Nanoscale Science and Technology, Rice Quantum Institute, and Department of Chemistry and Physics, MS-100, Rice University, P.O. Box 1892, Houston, TX 77251

[#]Institute for Theoretical Physics, University of California, Santa Barbara, CA 93106-4030

**An interacting one-dimensional (1D) electron system is predicted to behave very differently than its higher-dimensional counterparts.[1] Coulomb interactions strongly modify the properties away from those of a Fermi liquid, resulting in a Luttinger liquid (LL) characterized by a power-law vanishing of the density of states at the Fermi level. Experiments on one-dimensional semiconductor wires[2] and fractional quantum Hall conductors[3,4,5,6] have been interpreted using this picture, but questions remain about the connection between theory and experiment. Recently, single-walled carbon nanotubes (SWNTs) have emerged as a new type of 1D conductor[7,8,9,10] that may exhibit LL behavior[11,12]. Here we present measurements of the conductance of individual ropes of such SWNTs as a function of temperature and voltage. Power law behavior as a function of temperature or bias voltage is observed: $G \sim T^{\alpha}$ and $dI/dV \sim V^{\alpha}$. Both the power-law functional forms and the inferred exponents are in good agreement with theoretical predictions for tunneling into a LL.**

Since the initial discovery of SWNTs, experiments have revealed a great deal about their electronic properties. STM measurements of individual tubes have verified that they are either 1D semiconductors or conductors, depending upon their chirality[7,8]. Electrodes have also been attached to nanotubes and ropes of nanotubes to probe transport. These electrodes make tunneling contacts to the tubes and, for conducting tubes, the resulting structure behaves as a 1D quantum dot. It was found that for ropes transport was typically dominated by a single nanotube in the rope[9]. This is reasonable since the majority of the tubes comprising a rope are insulating at low temperatures[7,8]. Measurements of such devices have been used to study the charging energy, level spacing, and spin state of a nanotube[9,10,13,14].

The devices used in these previous experiments had two distinct geometries, once which contacted the ends of a tube and one which contacted the bulk, as is discussed in the legend to Fig. 1. Here, we explore the transport properties of rope samples in both of the geometries. Fig. 1 shows the linear-response two-terminal conductance, $G$, versus gate voltage, $V_g$, for a bulk-contacted metallic rope. At low temperatures, it exhibits a series of Coulomb oscillations[15] that occur each time that an electron is added to a nanotube within the rope. From the temperature dependence, we find that the charging energy $U$ for this sample is 1.9 meV. For $k_BT > U$ (i.e. $T > 20$ K), the Coulomb oscillations are nearly completely washed out, and the conductance is independent of gate voltage. A plot of the conductance vs. temperature in this regime is shown in the inset. The conductance drops steeply as the temperature is lowered, extrapolating to $G = 0$ at $T = 0$.

Results for a number of samples are shown in Fig. 2, where the $G$ versus $T$ is plotted on a log-log scale (solid lines). Fig. 2(a) shows data for end-contacted ropes, whereas Fig. 2(b) shows the data for bulk-contacted ropes. The measured data (solid lines) shows approximate power law behavior for the four samples shown. However, the range of temperature over which the power law behavior occurs is limited by the effects of Coulomb blockade at low temperatures. After correcting for the known temperature dependence due to the Coulomb blockade[15], the corrected data (dashed lines) shows power law behavior over a greater range, with slightly different exponents. Above T~100 K, G begins to saturate for some samples. This saturation is observed in many, but not all, of the samples studied.

Focusing on the corrected data, the bulk-contacted samples show approximate power law behavior from 8 – 300 K with exponents $\alpha_{bulk} \approx 0.33$ and 0.38. The end-contacted samples show approximate power law behavior from 10 - 100 K with exponents $\alpha_{end} \approx 0.6$ for both samples. The upper inset to Fig.

2(a) shows the exponents determined from the temperature dependence for a variety of samples. Exponents marked with an 'x' and 'o' are for bulk- and end- contacted tubes, respectively. The bulk-contacted samples show a systematically lower exponent than end-contacted devices, with $\alpha_{end} \sim 0.6$ and $\alpha_{bulk} \sim 0.3$.

Figure 3 shows the measured differential conductance $dI/dV$ of these devices as a function of the applied bias $V$. The upper left inset to Fig. 3(a) shows results for a bulk-contacted device at different temperatures, plotted on a log-log scale. At low biases, $dI/dV$ is proportional to a (temperature-dependent) constant – $G(T)$ from Figure 2. At high biases $dI/dV$ increases with increasing $V$. The curves at different temperatures fall onto a single curve in the high bias regime. Since this curve is roughly linear on a log-log plot, it implies that the differential conductance is described by a power law, $dI/dV \sim V^{\alpha}$, where $\alpha = 0.36$. At the lowest temperature $T=1.6$ K, this power law behavior occurs over two decades in $V$, from 1 mV $< V <$ 100 mV.

The upper left inset to Fig. 3(b) shows $dI/dV$ as a function of $V$ for an end-contacted sample at several temperatures. The conductance is again a temperature-dependent constant at low biases $eV \ll k_BT$, whereas at higher biases $dI/dV$ increases. The high bias data follows an approximate power law before rolling off to reduced slope for $V > 30$ mV. While the range of data is too small to conclude that a power law accurately describes the behavior at intermediate voltages, if a straight line is fit to the range 9 mV $< V <$ 32 mV the exponent obtained is $\alpha = 0.87$.

We now discuss the possible origins this approximate power law behavior. One possible explanation is that the tunnel barriers are strongly energy-dependent, with increased transparency at high energies. This would lead, e.g. to activated transport: $G \sim \exp(-\Delta/k_BT)$ over the barrier. However, the fact that the temperature dependence extrapolates to $G = 0$ at $T = 0$ (Figure 1 inset) is inconsistent with this functional form.

Another potential explanation is that transport occurs through multiple dots in series formed by disorder[16] or by barriers produced when the nanotubes bend over the lithographically defined contacts[17].

We rule this possibility out however, as we have chosen to study only samples where a single dominant period for the Coulomb oscillations is observed at low temperatures. This indicates the existence of only a single dot.

Having excluded these possibilities, let us consider whether the behavior can be explained by the predictions of Luttinger Liquid (LL) theory. A LL is a one-dimensional correlated electron state characterized by a parameter $g$ that measures the strength of the interaction between electrons. For any g $\neq$ 1, the low energy excitations of the system are not all weakly interacting quasiparticles, and the Fermi liquid theory used to describe conventional metals is not appropriate.

In SWNTs, the long-range Coulomb interaction between electrons is expected to yield an LL with $g < 1$ [11,12]. For a finite length tube or rope, the Luttinger parameter g is given by:

$$g = [1 + \frac{2U}{\Delta}]^{-1/2} \quad (1)$$

where $U = e^2/C$ is the charging energy of the tube and $\Delta = \pi \hbar v_F/2L$ is the single-particle level spacing (the two 1D subbands of the nanotube are assumed to be non-degenerate). From previous measurements and theoretical estimates[9,10] $U/\Delta \sim 6$., yielding an expected Luttinger parameter $g$(theory) $\sim 0.28$.

Tunneling of an electron into a LL is dramatically different than tunneling into Fermi liquid. For a Fermi liquid, an energy-independent tunneling amplitude is expected for energies near $E_F$, where $E_F$ is the Fermi level. This yields a temperature- and bias-independent tunneling conductance. For a clean LL on the other hand, the tunneling amplitude is predicted to vanish as a power law in $E-E_F$. This leads to a power-law variation of G with T at small biases $(eV << k_BT)$:

$$G(T) \sim AT^\alpha \quad (2)$$

or with V at large biases $(eV >> k_BT)$:

$$dI/dV \sim V^\alpha \quad (3).$$

The exponent of these power laws depends on the number of 1D channels[18] and whether the electron tunnels into the bulk or the end of the LL. For a SWNT with four conducting modes at $E_F$, the exponents are [11,12]:

$$\alpha_{end} = (g^{-1} - 1) / 4 \qquad (4a)$$

$$\alpha_{bulk} = (g^{-1} + g - 2) / 8. \qquad (4b)$$

Using equations 1 and 4, we obtain $\alpha_{end}$(theory) = 0.65 and $\alpha_{bulk}$(theory) = 0.24.

To compare the theoretical predictions for tunneling into an isolated nanotube through a single barrier to the experimental geometry where ropes are connected by two contacts, we must make two assumptions. First we assume that transport in the rope is dominated by a single metallic tube, as discussed previously. Preliminary theoretical studies[19] of ropes composed of SWNTs with a relatively small fraction of metallic tubes support this assumption. These studies find that the only significant inter-tube coupling is electrostatic. Such an interaction will introduce extra screening of the Coulomb interaction but, because of the weak (logarithmic) dependence of g on the screening length, the LL predictions are essentially unchanged. Second, we assume that the tunnel resistances into and out of the tube are the dominant resistances in the system. The circuit thus consists of two tunnel junctions in series, with the current response of each junction is described by equations 1-4. Note that the voltage drop across the highest impedance junction will be some fraction $\gamma$ of the total applied bias *V*, where $1/2 \leq \gamma \leq 1$. If the barriers are equal, the voltage will divide equally between these junctions and $\gamma = 1/2$. Alternately, if the resistance of one junction dominates, $\gamma = 1$.

With these assumptions, the approximate power law behavior as a function of *T* or *V* observed in Figures 2 and 3 then follows from Eq.s 1-4. The predicted values of the exponents are also in very good agreement with the experimental values. This agreement may be somewhat fortuitous due to the experimental uncertainty in the value of *U/Δ* and complexities associated with the screening of the Coulomb interaction by the metallic leads[11,12]. Nevertheless, the measurements are both qualitatively and quantitatively described by LL theory. Remarkably, power-law behavior in T is observed up to 300 K in the bulk-contacted samples, indicating that nanotubes are Luttinger Liquids even at room temperature.

At present, we do not understand the origins of the high-energy saturation observed in the end-contacted tubes. One possibility is that, at high energies, electrons can tunnel in both directions and hence the end-contacted tubes behave as bulk-contacted tubes, with a correspondingly lower exponent. Future experiments are necessary to clarify this issue.

The LL theory makes an additional prediction for this system. The differential conductance for a single tunnel junction is given by a universal scaling curve[20,21]:

$$\frac{dI}{dV} = AT^{\alpha} \cosh\left(\gamma \frac{eV}{2k_BT}\right) |\Gamma(\frac{1+\alpha}{2} + \gamma \frac{ieV}{2\pi k_BT})|^2 \qquad (5)$$

where $\Gamma(x)$ is the gamma function, $\gamma$ is the constant introduced earlier that takes into account the voltage division between the two tunnel junctions, and $A$ is an arbitrary constant. This equation assumes that the leads are at $T = 0$ K. For leads at a finite temperature, $dI/dV$ is given by the convolution of Eq. 5 and the derivative of the Fermi distribution: $df/dE = \frac{1}{4k_BT} \mathrm{sech}^2(\gamma eV/2k_BT)$.

If the above scaling relation is correct, it should be possible to collapse the data at different temperatures onto a single universal curve. To do this, the measured $dI/dV$ at each temperature was divided by $T^{\alpha}$ and plotted against $eV/k_BT$, as shown in the main body of Figs. 3(a) and 3(b). For both geometries, the scaled conductance is constant at as $eV/k_BT$ approaches zero, but above $eV/k_BT \sim 7$, the scaled curve begins to increase. The data collapses quite well onto a universal curve for the bulk-contacted device, Fig. 3(a), over the entire bias range. For the end-contacted device, the data deviates from power-law behavior for biases $V > 30$ mV as discussed previously. This is reflected in Fig. 3(b) in a roll-off that occurs at lower values of $eV/k_BT$ as the temperature is increased.

The solid lines in Fig. 3(a) and 3(b) are a plot of the curve obtained by fitting Eq. 5 (convolved with $df/dE$) to the data, with $\gamma$ as a fitting parameter. The theory fits the scaled data reasonably well, especially for the bulk-contacted tube. For the samples studied, the inferred values of $\gamma$ fall, within error bars, of the allowable range ($0.5 < \gamma < 1$) for two barriers in series. This indicates that energy scale at which the differential conductance makes the transition from a constant to power law behavior is well described by the theory.

Taken as a whole, the data shown in Fig. 2 and Fig. 3 provide strong evidence that the electrons in metallic carbon nanotubes constitute a Luttinger liquid. Future work will test other predictions of the LL theory, such as tunneling between LLs in end-to-end[1] and in crossed geometries[22].

**Acknowledgements**: We thank Steven Louie, Marvin Cohen, Dung-hai Lee, Alex Zettl, and Antoine Georges for useful discussions. This work was supported by DOE (Basic Energy Sciences, Materials Sciences Division, the sp$^2$ Materials Initiative). L. Balents was supported by the NSF. Correspondence should be addressed to P.L.M. (mceuen@socrates.berkeley.edu)

**Figures**

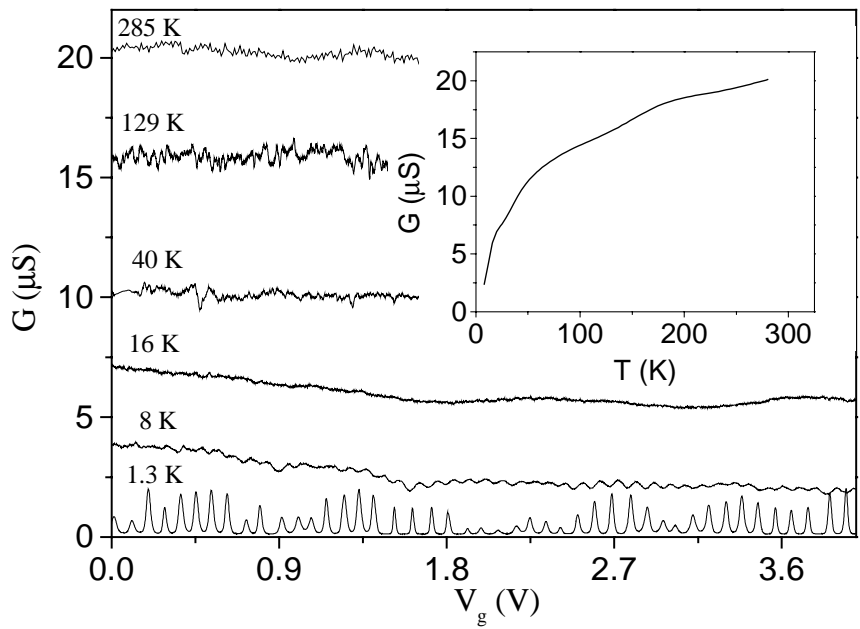

**Figure 1.** The two-terminal linear-response conductance G vs. gate voltage $V_g$ for a bulk-contacted metallic rope at a variety of temperatures. The data show significant temperature dependence for energy scales above the charging energy that cannot be explained by the Coulomb blockade model. Inset: Average conductance plotted as a function of temperature T. The devices used in these experiments are made in one of two ways. In both methods, SWNTs are deposited from a suspension in dichloroethane onto 1-µm thick $SiO_2$ that has been thermally grown on a degenerately doped Si wafer. The degenerately doped silicon substrate is used as a gate electrode. AFM imaging reveals that the diameter of the ropes vary between 1 and 10 nm. In the first method[9], chromium-gold contacts are applied over the top of the nanotube rope using electron beam lithography and lift-off. From measurements of these devices in the Coulomb blockade regime, we conclude that the electrons are confined to the length of rope between the leads. This implies that the leads cut the nanotubes into segments, and transport involves tunneling into the ends of the nanotubes ("end-contacted"). In the second method[10], electron beam lithography is first used to define leads, and ropes are deposited on top of the leads. Samples were selected that showed Coulomb blockade behavior at low temperatures with a single well-defined period, indicating the presence of a single dot. The charging energy of these samples indicates a dot with a size substantially

larger than the spacing between the leads, as found by Tans, et al.[10] Transport thus occurs by electrons tunneling into the middle, or bulk, of the nanotubes ("bulk-contacted").

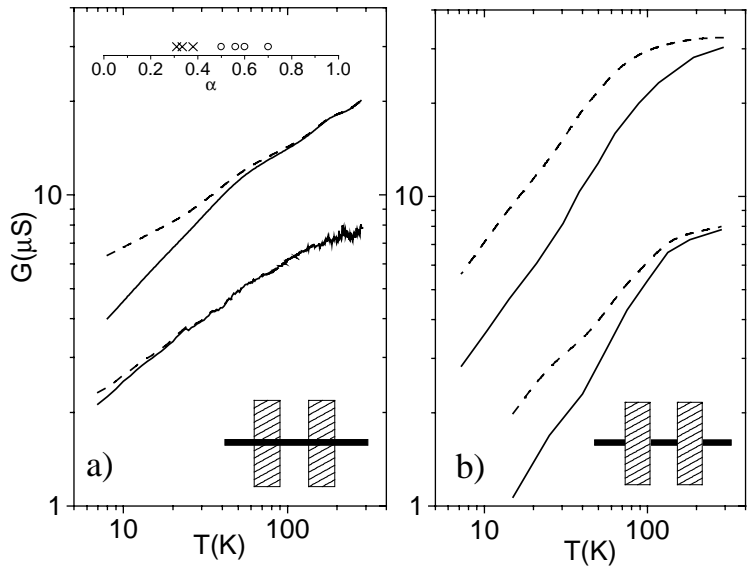

**Figure 2.** Conductance $G$ plotted against temperature $T$ for individual nanotube ropes. The data are plotted on a log-log scale. Figure 2a shows data for ropes that are deposited over pre-defined leads (bulk-contacted), whereas Fig. 2b shows the data for ropes that are contacted by evaporating the leads on top of the ropes (end-contacted). Sketches depicting the measurement configuration are shown in the lower right insets. The plots show both the bare data (solid line) and the data corrected for the temperature dependence expected from the Coulomb blockade (CB) model (dashed line). We correct the data by dividing the measured $G(T)$ by the theoretically expected temperature dependence in the CB model. This correction factor only depends upon $U/k_BT$, and, since $U$ can be independently measured from the temperature dependence of the Coulomb oscillations, the correction procedure requires no adjustable parameters. If the CB were the only source of the temperature dependence, the dashed lines would be horizontal. Instead they have a finite slope, indicating an approximate power-law dependence on $T$. The upper left inset to Figure 2(a) shows the power-law exponents inferred for a variety of samples. Open circles denote end-contacted devices, while crosses denote bulk-contacted ones.

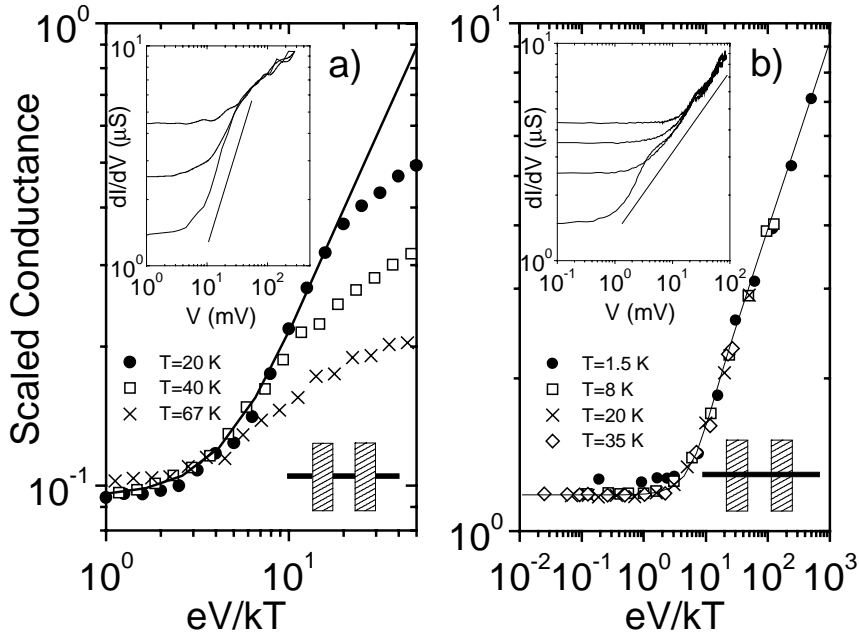

**Figure 3.** The differential conductance *dI/dV* measured at various temperatures. Figure 3a inset: *dI/dV* curves taken on a bulk-contacted rope at temperatures T=1.6 K, 8 K, 20 K, and 35K. Figure 3b inset: *dI/dV* curves taken on an end-contacted rope at temperatures T=20 K, 40 K, and 67K. In both insets, a straight line on the log-log plot is shown as a guide to the eye to indicate power-law behavior. The main panels show these measurements collapsed onto a single curve using the scaling relations described in the text. The solid line is the theoretical result fit to the data using $\gamma$ as a fitting parameter. The values of $\gamma$ resulting in the best fit to the data are $\gamma = 0.46$ in (a) and $\gamma = 0.63$ in (b).